
\def\unlock{\catcode`@=11} 
\def\lock{\catcode`@=12} 
\unlock
%
%
%
%
%

\font\fourteenrm=cmr10 scaled\magstep2
\font\twelverm=cmr10 scaled\magstep1
\font\ninerm=cmr9          \font\sixrm=cmr6

\font\fourteenbf=cmbx10 scaled\magstep2
\font\twelvebf=cmbx10 scaled\magstep1

\font\seventeeni=cmmi10 scaled\magstep3     \skewchar\seventeeni='177
\font\fourteeni=cmmi10 scaled\magstep2      \skewchar\fourteeni='177
\font\twelvei=cmmi10 scaled\magstep1        \skewchar\twelvei='177
\font\ninei=cmmi9                           \skewchar\ninei='177
\font\sixi=cmmi6                            \skewchar\sixi='177
\font\seventeensy=cmsy10 scaled\magstep3    \skewchar\seventeensy='60
\font\fourteensy=cmsy10 scaled\magstep2     \skewchar\fourteensy='60
\font\twelvesy=cmsy10 scaled\magstep1       \skewchar\twelvesy='60
\font\ninesy=cmsy9                          \skewchar\ninesy='60
\font\sixsy=cmsy6                           \skewchar\sixsy='60

\font\fourteenex=cmex10 scaled\magstep2
\font\twelveex=cmex10 scaled\magstep1

\font\fourteensl=cmsl10 scaled\magstep2
\font\twelvesl=cmsl10 scaled\magstep1

\font\fourteenit=cmti10 scaled\magstep2
\font\twelveit=cmti10 scaled\magstep1
\font\twelvett=cmtt10 scaled\magstep1
\font\twelvecp=cmcsc10 scaled\magstep1
\font\tencp=cmcsc10
\newfam\cpfam
%
%
\newcount\f@ntkey            \f@ntkey=0
\def\samef@nt{\relax \ifcase\f@ntkey \rm \or\oldstyle \or\or
         \or\it \or\sl \or\bf \or\tt \or\caps \fi }
\def\fourteenpoint{\relax
    \textfont0=\fourteenrm          \scriptfont0=\tenrm
    \scriptscriptfont0=\sevenrm
     \def\rm{\fam0 \fourteenrm \f@ntkey=0 }\relax
    \textfont1=\fourteeni           \scriptfont1=\teni
    \scriptscriptfont1=\seveni
     \def\oldstyle{\fam1 \fourteeni\f@ntkey=1 }\relax
    \textfont2=\fourteensy          \scriptfont2=\tensy
    \scriptscriptfont2=\sevensy
    \textfont3=\fourteenex     \scriptfont3=\fourteenex
    \scriptscriptfont3=\fourteenex
    \def\it{\fam\itfam \fourteenit\f@ntkey=4 }\textfont\itfam=\fourteenit
    \def\sl{\fam\slfam \fourteensl\f@ntkey=5 }\textfont\slfam=\fourteensl
    \scriptfont\slfam=\tensl
    \def\bf{\fam\bffam \fourteenbf\f@ntkey=6 }\textfont\bffam=\fourteenbf
    \scriptfont\bffam=\tenbf     \scriptscriptfont\bffam=\sevenbf
    \def\tt{\fam\ttfam \twelvett \f@ntkey=7 }\textfont\ttfam=\twelvett
    \h@big=11.9\p@{} \h@Big=16.1\p@{} \h@bigg=20.3\p@{} \h@Bigg=24.5\p@{}
    \def\caps{\fam\cpfam \twelvecp \f@ntkey=8 }\textfont\cpfam=\twelvecp
    \setbox\strutbox=\hbox{\vrule height 12pt depth 5pt width\z@}
    \samef@nt}
\def\twelvepoint{\relax
    \textfont0=\twelverm          \scriptfont0=\ninerm
    \scriptscriptfont0=\sixrm
     \def\rm{\fam0 \twelverm \f@ntkey=0 }\relax
    \textfont1=\twelvei           \scriptfont1=\ninei
    \scriptscriptfont1=\sixi
     \def\oldstyle{\fam1 \twelvei\f@ntkey=1 }\relax
    \textfont2=\twelvesy          \scriptfont2=\ninesy
    \scriptscriptfont2=\sixsy
    \textfont3=\twelveex          \scriptfont3=\twelveex
    \scriptscriptfont3=\twelveex
    \def\it{\fam\itfam \twelveit \f@ntkey=4 }\textfont\itfam=\twelveit
    \def\sl{\fam\slfam \twelvesl \f@ntkey=5 }\textfont\slfam=\twelvesl
    \scriptfont\slfam=\ninerm
    \def\bf{\fam\bffam \twelvebf \f@ntkey=6 }\textfont\bffam=\twelvebf
    \scriptfont\bffam=\ninerm     \scriptscriptfont\bffam=\sixrm
    \def\tt{\fam\ttfam \twelvett \f@ntkey=7 }\textfont\ttfam=\twelvett
    \h@big=10.2\p@{}
    \h@Big=13.8\p@{}
    \h@bigg=17.4\p@{}
    \h@Bigg=21.0\p@{}
    \def\caps{\fam\cpfam \twelvecp \f@ntkey=8 }\textfont\cpfam=\twelvecp
    \setbox\strutbox=\hbox{\vrule height 10pt depth 4pt width\z@}
    \samef@nt}
\def\tenpoint{\relax
    \textfont0=\tenrm          \scriptfont0=\sevenrm
    \scriptscriptfont0=\fiverm
    \def\rm{\fam0 \tenrm \f@ntkey=0 }\relax
    \textfont1=\teni           \scriptfont1=\seveni
    \scriptscriptfont1=\fivei
    \def\oldstyle{\fam1 \teni \f@ntkey=1 }\relax
    \textfont2=\tensy          \scriptfont2=\sevensy
    \scriptscriptfont2=\fivesy
    \textfont3=\tenex          \scriptfont3=\tenex
    \scriptscriptfont3=\tenex
    \def\it{\fam\itfam \tenit \f@ntkey=4 }\textfont\itfam=\tenit
    \def\sl{\fam\slfam \tensl \f@ntkey=5 }\textfont\slfam=\tensl
    \def\bf{\fam\bffam \tenbf \f@ntkey=6 }\textfont\bffam=\tenbf
    \scriptfont\bffam=\sevenbf     \scriptscriptfont\bffam=\fivebf
    \def\tt{\fam\ttfam \tentt \f@ntkey=7 }\textfont\ttfam=\tentt
    \def\caps{\fam\cpfam \tencp \f@ntkey=8 }\textfont\cpfam=\tencp
    \setbox\strutbox=\hbox{\vrule height 8.5pt depth 3.5pt width\z@}
    \samef@nt}
%
%
%
%
\newdimen\h@big  \h@big=8.5\p@
\newdimen\h@Big  \h@Big=11.5\p@
\newdimen\h@bigg  \h@bigg=14.5\p@
\newdimen\h@Bigg  \h@Bigg=17.5\p@
\def\big#1{{\hbox{$\left#1\vbox to\h@big{}\right.\n@space$}}}
\def\Big#1{{\hbox{$\left#1\vbox to\h@Big{}\right.\n@space$}}}
\def\bigg#1{{\hbox{$\left#1\vbox to\h@bigg{}\right.\n@space$}}}
\def\Bigg#1{{\hbox{$\left#1\vbox to\h@Bigg{}\right.\n@space$}}}
%
%
%
\normalbaselineskip = 20pt plus 0.2pt minus 0.1pt
\normallineskip = 1.5pt plus 0.1pt minus 0.1pt
\normallineskiplimit = 1.5pt
\newskip\normaldisplayskip
\normaldisplayskip = 18pt plus 4pt minus 8pt
\newskip\normaldispshortskip
\normaldispshortskip = 5pt plus 4pt
\newskip\normalparskip
\normalparskip = 6pt plus 2pt minus 1pt
\newskip\skipregister
\skipregister = 5pt plus 2pt minus 1.5pt
\newif\ifsingl@    \newif\ifdoubl@
\newif\iftwelv@    \twelv@true
\def\singlespace{\singl@true\doubl@false\spaces@t}
\def\doublespace{\singl@false\doubl@true\spaces@t}
\def\normalspace{\singl@false\doubl@false\spaces@t}
\def\Tenpoint{\tenpoint\twelv@false\spaces@t}
\def\Twelvepoint{\twelvepoint\twelv@true\spaces@t}
\def\spaces@t{\relax%
 \iftwelv@ \ifsingl@\subspaces@t3:4;\else\subspaces@t1:1;\fi%
 \else \ifsingl@\subspaces@t3:5;\else\subspaces@t4:5;\fi \fi%
 \ifdoubl@ \multiply\baselineskip by 5%
 \divide\baselineskip by 4 \fi \unskip}
\def\subspaces@t#1:#2;{%
      \baselineskip = \normalbaselineskip%
      \multiply\baselineskip by #1 \divide\baselineskip by #2%
      \lineskip = \normallineskip%
      \multiply\lineskip by #1 \divide\lineskip by #2%
      \lineskiplimit = \normallineskiplimit%
      \multiply\lineskiplimit by #1 \divide\lineskiplimit by #2%
      \parskip = \normalparskip%
      \multiply\parskip by #1 \divide\parskip by #2%
      \abovedisplayskip = \normaldisplayskip%
      \multiply\abovedisplayskip by #1 \divide\abovedisplayskip by #2%
      \belowdisplayskip = \abovedisplayskip%
      \abovedisplayshortskip = \normaldispshortskip%
      \multiply\abovedisplayshortskip by #1%
        \divide\abovedisplayshortskip by #2%
      \belowdisplayshortskip = \abovedisplayshortskip%
      \advance\belowdisplayshortskip by \belowdisplayskip%
      \divide\belowdisplayshortskip by 2%
      \smallskipamount = \skipregister%
      \multiply\smallskipamount by #1 \divide\smallskipamount by #2%
      \medskipamount = \smallskipamount \multiply\medskipamount by 2%
      \bigskipamount = \smallskipamount \multiply\bigskipamount by 4 }
\def\normalbaselines{ \baselineskip=\normalbaselineskip%
   \lineskip=\normallineskip \lineskiplimit=\normallineskip%
   \iftwelv@\else \multiply\baselineskip by 4 \divide\baselineskip by 5%
     \multiply\lineskiplimit by 4 \divide\lineskiplimit by 5%
     \multiply\lineskip by 4 \divide\lineskip by 5 \fi }
\Twelvepoint  
\interlinepenalty=50
\interfootnotelinepenalty=5000
\predisplaypenalty=9000
\postdisplaypenalty=500
\hfuzz=1pt
\vfuzz=0.2pt
%
%
%
\def\pagecontents{%
   \ifvoid\topins\else\unvbox\topins\vskip\skip\topins\fi
   \dimen@ = \dp255 \unvbox255
   \ifvoid\footins\else\vskip\skip\footins\footrule\unvbox\footins\fi
   \ifr@ggedbottom \kern-\dimen@ \vfil \fi }
\def\makeheadline{\vbox to 0pt{ \skip@=\topskip
      \advance\skip@ by -12pt \advance\skip@ by -2\normalbaselineskip
      \vskip\skip@ \line{\vbox to 12pt{}\the\headline} \vss
      }\nointerlineskip}
\def\makefootline{\baselineskip = 1.5\normalbaselineskip
                 \line{\the\footline}}
\newif\iffrontpage
\newif\ifletterstyle
\newif\ifp@genum
\def\nopagenumbers{\p@genumfalse}
\def\pagenumbers{\p@genumtrue}
\pagenumbers
\newtoks\paperheadline
\newtoks\letterheadline
\newtoks\letterfrontheadline
\newtoks\lettermainheadline
\newtoks\paperfootline
\newtoks\letterfootline
\newtoks\date
\footline={\ifletterstyle\the\letterfootline\else\the\paperfootline\fi}
\paperfootline={\hss\iffrontpage\else\ifp@genum\tenrm\folio\hss\fi\fi}
\letterfootline={\hfil}
\headline={\ifletterstyle\the\letterheadline\else\the\paperheadline\fi}
\paperheadline={\hfil}
\letterheadline{\iffrontpage\the\letterfrontheadline
     \else\the\lettermainheadline\fi}
\lettermainheadline={\rm\ifp@genum page \ \folio\fi\hfil\the\date}
\def\monthname{\relax\ifcase\month 0/\or January\or February\or
   March\or April\or May\or June\or July\or August\or September\or
   October\or November\or December\else\number\month/\fi}
\date={\monthname\ \number\day, \number\year}
\countdef\pagenumber=1  \pagenumber=1
\def\advancepageno{\global\advance\pageno by 1
   \ifnum\pagenumber<0 \global\advance\pagenumber by -1
    \else\global\advance\pagenumber by 1 \fi \global\frontpagefalse }
\def\folio{\ifnum\pagenumber<0 \romannumeral-\pagenumber
           \else \number\pagenumber \fi }
\def\footrule{\dimen@=\prevdepth\nointerlineskip
   \vbox to 0pt{\vskip -0.25\baselineskip \hrule width 0.35\hsize \vss}
   \prevdepth=\dimen@ }
\newtoks\foottokens
\foottokens={\Tenpoint\singlespace}
\newdimen\footindent
\footindent=24pt
\def\vfootnote#1{\insert\footins\bgroup  \the\foottokens
   \interlinepenalty=\interfootnotelinepenalty \floatingpenalty=20000
   \splittopskip=\ht\strutbox \boxmaxdepth=\dp\strutbox
   \leftskip=\footindent \rightskip=\z@skip
   \parindent=0.5\footindent \parfillskip=0pt plus 1fil
   \spaceskip=\z@skip \xspaceskip=\z@skip
   \Textindent{$ #1 $}\footstrut\futurelet\next\fo@t}
\def\Textindent#1{\noindent\llap{#1\enspace}\ignorespaces}
\def\footnote#1{\attach{#1}\vfootnote{#1}}

\let\footsymbol=\star
\newcount\lastf@@t           \lastf@@t=-1
\newcount\footsymbolcount    \footsymbolcount=0
\newif\ifPhysRev
\def\footsymbolgen{\relax \ifPhysRev \iffrontpage \NPsymbolgen\else
      \PRsymbolgen\fi \else \NPsymbolgen\fi
   \global\lastf@@t=\pageno \footsymbol }
\def\NPsymbolgen{\ifnum\footsymbolcount<0 \global\footsymbolcount=0\fi
   {\iffrontpage \else \advance\lastf@@t by 1 \fi
    \ifnum\lastf@@t<\pageno \global\footsymbolcount=0
     \else \global\advance\footsymbolcount by 1 \fi }
   \ifcase\footsymbolcount \fd@f\star\or \fd@f\dagger\or \fd@f\ddagger\or
    \fd@f\ast\or \fd@f\natural\or \fd@f\diamond\or \fd@f\bullet\or
    \fd@f\nabla\else \fd@f\dagger\global\footsymbolcount=0 \fi }
\def\fd@f#1{\xdef\footsymbol{#1}}
\def\PRsymbolgen{\ifnum\footsymbolcount>0 \global\footsymbolcount=0\fi
      \global\advance\footsymbolcount by -1
      \xdef\footsymbol{\sharp\number-\footsymbolcount} }
\def\space@ver#1{\let\@sf=\empty \ifmmode #1\else \ifhmode
   \edef\@sf{\spacefactor=\the\spacefactor}\unskip${}#1$\relax\fi\fi}
\def\attach#1{\space@ver{\strut^{\mkern 2mu #1} }\@sf\ }
%
%
%
\newcount\chapternumber      \chapternumber=0
\newcount\sectionnumber      \sectionnumber=0
\newcount\equanumber         \equanumber=0
\let\chapterlabel=0
\newtoks\chapterstyle        \chapterstyle={\Number}
\newskip\chapterskip         \chapterskip=\bigskipamount
\newskip\sectionskip         \sectionskip=\medskipamount
\newskip\headskip            \headskip=8pt plus 3pt minus 3pt
\newdimen\chapterminspace    \chapterminspace=15pc
\newdimen\sectionminspace    \sectionminspace=10pc
\newdimen\referenceminspace  \referenceminspace=25pc
\def\chapterreset{\global\advance\chapternumber by 1
   \ifnum\equanumber<0 \else\global\equanumber=0\fi
   \sectionnumber=0 \makel@bel}
\def\makel@bel{\xdef\chapterlabel{%
\the\chapterstyle{\the\chapternumber}.}}
\def\sectionlabel{\number\sectionnumber \quad }
\def\alphabetic#1{\count255='140 \advance\count255 by #1\char\count255}
\def\Alphabetic#1{\count255='100 \advance\count255 by #1\char\count255}
\def\Roman#1{\uppercase\expandafter{\romannumeral #1}}
\def\roman#1{\romannumeral #1}
\def\Number#1{\number #1}
\def\unnumberedchapters{\let\makel@bel=\relax \let\chapterlabel=\relax
\let\sectionlabel=\relax \equanumber=-1 }
\def\titlestyle#1{\par\begingroup \interlinepenalty=9999
     \leftskip=0.02\hsize plus 0.23\hsize minus 0.02\hsize
     \rightskip=\leftskip \parfillskip=0pt
     \hyphenpenalty=9000 \exhyphenpenalty=9000
     \tolerance=9999 \pretolerance=9000
     \spaceskip=0.333em \xspaceskip=0.5em
     \iftwelv@\twelvepoint\fourteenrm\else\twelvepoint\fi
   \noindent #1\par\endgroup }
\def\spacecheck#1{\dimen@=\pagegoal\advance\dimen@ by -\pagetotal
   \ifdim\dimen@<#1 \ifdim\dimen@>0pt \vfil\break \fi\fi}
\def\chapter#1{\par \penalty-300 \vskip\chapterskip
   \spacecheck\chapterminspace
   \chapterreset \titlestyle{\chapterlabel \ #1}
   \nobreak\vskip\headskip \penalty 30000
   \wlog{\string\chapter\ \chapterlabel} }

\def\section#1{\par \ifnum\the\lastpenalty=30000\else
   \penalty-200\vskip\sectionskip \spacecheck\sectionminspace\fi
   \wlog{\string\section\ \chapterlabel \the\sectionnumber}
   \global\advance\sectionnumber by 1  \noindent
   {\caps\enspace\chapterlabel \sectionlabel #1}\par
   \nobreak\vskip\headskip \penalty 30000 }
\def\subsection#1{\par
   \ifnum\the\lastpenalty=30000\else \penalty-100\smallskip \fi
   \noindent\undertext{#1}\enspace \vadjust{\penalty5000}}

\def\undertext#1{\vtop{\hbox{#1}\kern 1pt \hrule}}
\def\APPENDIX#1#2{\par\penalty-300\vskip\chapterskip
   \spacecheck\chapterminspace \chapterreset \xdef\chapterlabel{#1}
   \titlestyle{APPENDIX #2} \nobreak\vskip\headskip \penalty 30000
   \wlog{\string\Appendix\ \chapterlabel} }
\def\Appendix#1{\APPENDIX{#1}{#1}}
\def\appendix{\APPENDIX{A}{}}
%
%
%

\newif\ifdraftmode
\draftmodefalse

\def\eqname#1{\relax \ifnum\equanumber<0
     \xdef#1{{(\number-\equanumber)}}\global\advance\equanumber by -1
    \else \global\advance\equanumber by 1
      \xdef#1{{(\chapterlabel \number\equanumber)}} \fi}
%



%
\def\eqinsert#1{\noalign{\dimen@=\prevdepth \nointerlineskip
   \setbox0=\hbox to\displaywidth{\hfil #1}
   \vbox to 0pt{\vss\hbox{$\!\box0\!$}\kern-0.5\baselineskip}
   \prevdepth=\dimen@}}
\def\sequentialequations{\equanumber=-1}
%
%
\def\GENITEM#1;#2{\par \hangafter=0 \hangindent=#1
    \Textindent{$ #2 $}\ignorespaces}
\outer\def\newitem#1=#2;{\gdef#1{\GENITEM #2;}}
\newdimen\itemsize                \itemsize=30pt
\newitem\item=1\itemsize;
\newitem\sitem=1.75\itemsize;     
\newitem\ssitem=2.5\itemsize;     
\outer\def\newlist#1=#2&#3&#4;{\toks0={#2}\toks1={#3}%
   \count255=\escapechar \escapechar=-1
   \alloc@0\list\countdef\insc@unt\listcount     \listcount=0
   \edef#1{\par
      \countdef\listcount=\the\allocationnumber
      \advance\listcount by 1
      \hangafter=0 \hangindent=#4
      \Textindent{\the\toks0{\listcount}\the\toks1}}
   \expandafter\expandafter\expandafter
    \edef\c@t#1{begin}{\par
      \countdef\listcount=\the\allocationnumber \listcount=1
      \hangafter=0 \hangindent=#4
      \Textindent{\the\toks0{\listcount}\the\toks1}}
   \expandafter\expandafter\expandafter
    \edef\c@t#1{con}{\par \hangafter=0 \hangindent=#4 \noindent}
   \escapechar=\count255}
\def\c@t#1#2{\csname\string#1#2\endcsname}
\newlist\point=\Number&.&1.0\itemsize;
\newlist\subpoint=(\alphabetic&)&1.75\itemsize;
\newlist\subsubpoint=(\roman&)&2.5\itemsize;
\let\spoint=\subpoint

%
%
%
\newcount\referencecount     \referencecount=0
\newif\ifreferenceopen       \newwrite\referencewrite
\newtoks\rw@toks
\def\NPrefmark#1{\attach{\scriptscriptstyle [ #1 ] }}
\let\PRrefmark=\attach
\def\refmark#1{\relax\ifPhysRev\PRrefmark{#1}\else\NPrefmark{#1}\fi}
\def\refend{\refmark{\number\referencecount}}
\newcount\lastrefsbegincount \lastrefsbegincount=0
\def\refsend{\refmark{\count255=\referencecount
   \advance\count255 by-\lastrefsbegincount
   \ifcase\count255 \number\referencecount
   \or \number\lastrefsbegincount,\number\referencecount
   \else \number\lastrefsbegincount-\number\referencecount \fi}}
\def\refch@ck{\chardef\rw@write=\referencewrite
   \ifreferenceopen \else \referenceopentrue
   \immediate\openout\referencewrite=referenc.tex \fi}
%
{\catcode`\^^M=\active 
  \gdef\obeyendofline{\catcode`\^^M\active \let^^M\ }}%
%
{\catcode`\^^M=\active 
  \gdef\ignoreendofline{\catcode`\^^M=5}}
{\obeyendofline\gdef\rw@start#1{\def\t@st{#1} \ifx\t@st\blankend%
\endgroup \@sf \relax \else \ifx\t@st\bl@nkend \endgroup \@sf \relax%
\else \rw@begin#1
\backtotext
\fi \fi } }
{\obeyendofline\gdef\rw@begin#1
{\def\n@xt{#1}\rw@toks={#1}\relax%
\rw@next}}
\def\blankend{}
{\obeylines\gdef\bl@nkend{
}}
\newif\iffirstrefline  \firstreflinetrue
\def\rwr@teswitch{\ifx\n@xt\blankend \let\n@xt=\rw@begin %
 \else\iffirstrefline \global\firstreflinefalse%
\immediate\write\rw@write{\noexpand\obeyendofline \the\rw@toks}%
\let\n@xt=\rw@begin%
      \else\ifx\n@xt\rw@@d \def\n@xt{\immediate\write\rw@write{%
        \noexpand\ignoreendofline}\endgroup \@sf}%
             \else \immediate\write\rw@write{\the\rw@toks}%
             \let\n@xt=\rw@begin\fi\fi \fi}
\def\rw@next{\rwr@teswitch\n@xt}
\def\rw@@d{\backtotext} \let\rw@end=\relax
\let\backtotext=\relax

\newdimen\refindent     \refindent=30pt
\def\refitem#1{\par \hangafter=0 \hangindent=\refindent \Textindent{#1}}
\def\REFNUM#1{\space@ver{}\refch@ck \firstreflinetrue%
 \global\advance\referencecount by 1 \xdef#1{\the\referencecount}}
\def\refnum#1{\space@ver{}\refch@ck \firstreflinetrue%
 \global\advance\referencecount by 1 \xdef#1{\the\referencecount}\refend}

\def\REF#1{\REFNUM#1%
 \immediate\write\referencewrite{%
            \noexpand\refitem{\ifdraftmode{\sevenrm
               \noexpand\string\string#1\ }\fi#1.}}%
      \begingroup\obeyendofline\rw@start}
\def\ref{\refnum\?%
 \immediate\write\referencewrite{\noexpand\refitem{\?.}}%
\begingroup\obeyendofline\rw@start}
\def\Ref#1{\refnum#1%
 \immediate\write\referencewrite{\noexpand\refitem{#1.}}%
\begingroup\obeyendofline\rw@start}
\def\REFS#1{\REFNUM#1\global\lastrefsbegincount=\referencecount
\immediate\write\referencewrite{\noexpand\refitem{#1.}}%
\begingroup\obeyendofline\rw@start}
%

%
%

%
\newcount\figurecount     \figurecount=0
\newif\iffigureopen       \newwrite\figurewrite
\def\figch@ck{\chardef\rw@write=\figurewrite \iffigureopen\else
   \immediate\openout\figurewrite=figures.aux
   \figureopentrue\fi}
\def\FIGNUM#1{\space@ver{}\figch@ck \firstreflinetrue%
 \global\advance\figurecount by 1 \xdef#1{\the\figurecount}}
\def\FIG#1{\FIGNUM#1
   \immediate\write\figurewrite{\noexpand\refitem{#1.}}%
   \begingroup\obeyendofline\rw@start}
\def\fig{\FIGNUM\? fig.~\?%
\immediate\write\figurewrite{\noexpand\refitem{\?.}}%
\begingroup\obeyendofline\rw@start}
\def\figure{\FIGNUM\? figure~\?
   \immediate\write\figurewrite{\noexpand\refitem{\?.}}%
   \begingroup\obeyendofline\rw@start}
\def\Fig{\FIGNUM\? Fig.~\?%
\immediate\write\figurewrite{\noexpand\refitem{\?.}}%
\begingroup\obeyendofline\rw@start}
\def\Figure{\FIGNUM\? Figure~\?%
\immediate\write\figurewrite{\noexpand\refitem{\?.}}%
\begingroup\obeyendofline\rw@start}
\newcount\tablecount     \tablecount=0
\newif\iftableopen       \newwrite\tablewrite
\def\tabch@ck{\chardef\rw@write=\tablewrite \iftableopen\else
   \immediate\openout\tablewrite=tables.aux
   \tableopentrue\fi}
\def\TABNUM#1{\space@ver{}\tabch@ck \firstreflinetrue%
 \global\advance\tablecount by 1 \xdef#1{\the\tablecount}}
\def\TABLE#1{\TABNUM#1
   \immediate\write\tablewrite{\noexpand\refitem{#1.}}%
   \begingroup\obeyendofline\rw@start}
\def\Table{\TABNUM\? Table~\?%
\immediate\write\tablewrite{\noexpand\refitem{\?.}}%
\begingroup\obeyendofline\rw@start}
\newif\ifsymbolopen       \newwrite\symbolwrite
\def\symch@ck{\ifsymbolopen\else
   \immediate\openout\symbolwrite=symbols.aux
   \symbolopentrue\fi}
\def\symdef#1#2{\def#1{#2}%
      \symch@ck%
      \immediate\write\symbolwrite{$$ \hbox{\noexpand\string\string#1}
                \noexpand\qquad\noexpand\longrightarrow\noexpand\qquad
                           \string#1 $$}}
%
%

%
%
%
\def\masterreset{\global\pagenumber=1 \global\chapternumber=0
   \global\equanumber=0 \global\sectionnumber=0
   \global\referencecount=0 \global\figurecount=0 \global\tablecount=0 }
\def\FRONTPAGE{\ifvoid255\else\vfill\penalty-2000\fi
      \masterreset\global\frontpagetrue
      \global\lastf@@t=0 \global\footsymbolcount=0}

\def\paperstyle{\letterstylefalse\normalspace\papersize}
\def\letterstyle{\letterstyletrue\singlespace\lettersize}
\def\papersize{\hsize=35pc\vsize=50pc\hoffset=1pc\voffset=6pc
               \skip\footins=\bigskipamount}
\def\lettersize{\hsize=6.5in\vsize=8.5in\hoffset=0in\voffset=1in
   \skip\footins=\smallskipamount \multiply\skip\footins by 3 }
\paperstyle   
%
%
\def\MEMO{\letterstyle\FRONTPAGE \letterfrontheadline={\hfil}
    \line{\quad\fourteenrm NTC MEMORANDUM\hfil\twelverm\the\date\quad}
    \medskip \memod@f}

\def\memit@m#1{\smallskip \hangafter=0 \hangindent=1in
      \Textindent{\caps #1}}
\def\memod@f{\xdef\to{\memit@m{To:}}\xdef\from{\memit@m{From:}}%
     \xdef\topic{\memit@m{Topic:}}\xdef\subject{\memit@m{Subject:}}%
     \xdef\rule{\bigskip\hrule height 1pt\bigskip}}
\memod@f
\newskip\lettertopfil
\lettertopfil = 0pt plus 1.5in minus 0pt
\newskip\letterbottomfil
\letterbottomfil = 0pt plus 2.3in minus 0pt
\newskip\spskip \setbox0\hbox{\ } \spskip=-1\wd0
\def\addressee#1{\medskip\rightline{\the\date\hskip 30pt} \bigskip
   \vskip\lettertopfil
   \ialign to\hsize{\strut ##\hfil\tabskip 0pt plus \hsize \cr #1\crcr}
   \medskip\noindent\hskip\spskip}
\newskip\signatureskip       \signatureskip=40pt
\def\signed#1{\par \penalty 9000 \bigskip \dt@pfalse
  \everycr={\noalign{\ifdt@p\vskip\signatureskip\global\dt@pfalse\fi}}
  \setbox0=\vbox{\singlespace \halign{\tabskip 0pt \strut ##\hfil\cr
   \noalign{\global\dt@ptrue}#1\crcr}}
  \line{\hskip 0.5\hsize minus 0.5\hsize \box0\hfil} \medskip }

\def\endletter{\ifnum\pagenumber=1 \vskip\letterbottomfil\supereject
\else \vfil\supereject \fi}
\newbox\letterb@x
\def\lettertext{\par\unvcopy\letterb@x\par}
\def\multiletter{\setbox\letterb@x=\vbox\bgroup
      \everypar{\vrule height 1\baselineskip depth 0pt width 0pt }
      \singlespace \topskip=\baselineskip }
\def\letterend{\par\egroup}
%
%
%
\newskip\frontpageskip
\newtoks\pubtype
\newtoks\Pubnum
\newtoks\pubnum
\newif\ifp@bblock  \p@bblocktrue
\def\PH@SR@V{\doubl@true \baselineskip=24.1pt plus 0.2pt minus 0.1pt
             \parskip= 3pt plus 2pt minus 1pt }
\def\PHYSREV{\paperstyle\PhysRevtrue\PH@SR@V}
\def\titlepage{\FRONTPAGE\paperstyle\ifPhysRev\PH@SR@V\fi
   \ifp@bblock\p@bblock\fi}
\def\nopubblock{\p@bblockfalse}
\def\endpage{\vfil\break}
\frontpageskip=1\medskipamount plus .5fil
\pubtype={\tensl Preliminary Version}
\Pubnum={$\caps SLAC - PUB - \the\pubnum $}
\pubnum={0000}
\def\p@bblock{\begingroup \tabskip=\hsize minus \hsize
   \baselineskip=1.5\ht\strutbox \topspace-2\baselineskip
   \halign to\hsize{\strut ##\hfil\tabskip=0pt\crcr
   \the\Pubnum\cr \the\date\cr \the\pubtype\cr}\endgroup}
\def\title#1{\vskip\frontpageskip \titlestyle{#1} \vskip\headskip }
\def\author#1{\vskip\frontpageskip\titlestyle{\twelvecp #1}\nobreak}

\def\address#1{\par\kern 5pt\titlestyle{\twelvepoint\it #1}}
\def\andaddress{\par\kern 5pt \centerline{\sl and} \address}

\def\abstract{\vskip\frontpageskip\centerline{\fourteenrm ABSTRACT}
              \vskip\headskip }

%
%
%

\def\\{\relax\ifmmode\backslash\else$\backslash$\fi}
\def\globaleqnumbers{\relax\if\equanumber<0\else\global\equanumber=-1\fi}

\def\journal#1&#2(#3){\unskip, \sl #1~\bf #2 \rm (19#3) }

\def\topspace{\hrule height 0pt depth 0pt \vskip}

\let\int=\intop         
\def\prop{\mathrel{{\mathchoice{\pr@p\scriptstyle}{\pr@p\scriptstyle}{
                \pr@p\scriptscriptstyle}{\pr@p\scriptscriptstyle} }}}
\def\pr@p#1{\setbox0=\hbox{$\cal #1 \char'103$}
   \hbox{$\cal #1 \char'117$\kern-.4\wd0\box0}}
\def\lsim{\mathrel{\mathpalette\@versim<}}
\def\gsim{\mathrel{\mathpalette\@versim>}}
\def\@versim#1#2{\lower0.5ex\vbox{\baselineskip\z@skip\lineskip-.1ex
  \lineskiplimit\z@\ialign{$\m@th#1\hfil##\hfil$\crcr#2\crcr\sim\crcr}}}
%
%
%
\let\sec@nt=\sec
\def\sec{\relax\ifmmode\let\n@xt=\sec@nt\else\let\n@xt\section\fi\n@xt}
\def\obsolete#1{\message{Macro \string #1 is obsolete.}}
\def\firstsec#1{\obsolete\firstsec \section{#1}}
\def\firstsubsec#1{\obsolete\firstsubsec \subsection{#1}}
\def\thispage#1{\obsolete\thispage \global\pagenumber=#1\frontpagefalse}
\def\thischapter#1{\obsolete\thischapter \global\chapternumber=#1}
\def\nextequation#1{\obsolete\nextequation \global\equanumber=#1
   \ifnum\the\equanumber>0 \global\advance\equanumber by 1 \fi}
\def\BOXITEM{\afterassigment\B@XITEM\setbox0=}
\def\B@XITEM{\par\hangindent\wd0 \noindent\box0 }
%

%
%
%
%
%
\lock
\message{   }
%
%
%












\newbox\figbox
\newdimen\zero  \zero=0pt
\newdimen\figmove
\newdimen\figwidth
\newdimen\figheight
\newdimen\textwidth
\newtoks\figtoks
\newcount\figcounta
\newcount\figcountb
\newcount\figlines
\def\figreset{\global\figcounta=-1 \global\figcountb=-1
\global\figmove=\baselineskip
\global\figlines=1 \global\figtoks={ } }
\def\picture#1by#2:#3{\global\setbox\figbox=\vbox{\vskip #1
\hbox{\vbox{\hsize=#2 \noindent #3}}}
\global\setbox\figbox=\vbox{\kern 10pt
\hbox{\kern 10pt \box\figbox \kern 10pt }\kern 10pt}
\global\figwidth=1\wd\figbox
\global\figheight=1\ht\figbox
\global\textwidth=\hsize
\global\advance\textwidth by - \figwidth }
\def\figtoksappend{\edef\temp##1{\global\figtoks=%
{\the\figtoks ##1}}\temp}
\def\figparmsa#1{\loop \global\advance\figcounta by 1
\ifnum \figcounta < #1
\figtoksappend{ 0pt \the\hsize }
\global\advance\figlines by 1
\repeat }
\def\figparmsb#1{\loop \global\advance\figcountb by 1
\ifnum \figcountb < #1
\figtoksappend{ \the\figwidth \the\textwidth}
\global\advance\figlines by 1
\repeat }
\def\figtext#1:#2:#3{\figreset%
\figparmsa{#1}%
\figparmsb{#2}%
\multiply\figmove by #1%
\global\setbox\figbox=\vbox to 0pt{\vskip \figmove  \hbox{\box\figbox}
\vss }
\parshape=\the\figlines\the\figtoks\the\zero\the\hsize
\noindent
\rlap{\box\figbox} #3}
\def\Buildrel#1\under#2{\mathrel{\mathop{#2}\limits_{#1}}}
\def\llongrarrow{\hbox to 40pt{\rightarrowfill}}



\def\boxit#1{\vbox{\hrule\hbox{\vrule\kern3pt
\vbox{\kern3pt#1\kern3pt}\kern3pt\vrule}\hrule}}
\newdimen\str
\def\fboxit#1#2{\vbox{\hrule height #1 \hbox{\vrule width #1
\kern3pt \vbox{\kern3pt#2\kern3pt}\kern3pt \vrule width #1 }
\hrule height #1 }}
\def\tran#1#2{\transpoint \hfuzz 5pt \gdef\label{#1}
\vbox to \the\vsize{\hsize \the\hsize #2} \par \eject }
\newdimen\baseskip
\newdimen\lskip
\lskip=\lineskip
\newdimen\transize
\newdimen\tall
\def\transpoint{\gdef\rm{\fam0\eighteenrm}
\font\twentyfourit = cmti10 scaled \magstep5
\font\twentyfourrm = cmr10 scaled \magstep5
\font\twentyfourbf = cmbx10 scaled \magstep5
\font\twentyeightsy = cmsy10 scaled \magstep5
\font\eighteenrm = cmr10 scaled \magstep3
\font\eighteenb = cmbx10 scaled \magstep3
\font\eighteeni = cmmi10 scaled \magstep3
\font\eighteenit = cmti10 scaled \magstep3
\font\eighteensl = cmsl10 scaled \magstep3
\font\eighteensy = cmsy10 scaled \magstep3
\font\eighteencaps = cmr10 scaled \magstep3
\font\eighteenmathex = cmex10 scaled \magstep3
\font\fourteenrm=cmr10 scaled \magstep2
\font\fourteeni=cmmi10 scaled \magstep2
\font\fourteenit = cmti10 scaled \magstep2
\font\fourteensy=cmsy10 scaled \magstep2
\font\fourteenmathex = cmex10 scaled \magstep2
\parindent 20pt
\global\hsize = 7.0in
\global\vsize = 8.9in
\dimen\transize = \the\hsize
\dimen\tall = \the\vsize
\def\sy{\eighteensy }
\def\sl{\eighteens }
\def\bf{\eighteenb }
\def\it{\eighteenit }
\def\caps{\eighteencaps }
\textfont0=\eighteenrm \scriptfont0=\fourteenrm
\scriptscriptfont0=\twelverm
\textfont1=\eighteeni \scriptfont1=\fourteeni \scriptscriptfont1=\twelvei
\textfont2=\eighteensy \scriptfont2=\fourteensy
\scriptscriptfont2=\twelvesy
\textfont3=\eighteenmathex \scriptfont3=\eighteenmathex
\scriptscriptfont3=\eighteenmathex
\global\baselineskip 35pt
\global\lineskip 15pt
\global\parskip 5pt  plus 1pt minus 1pt
\global\abovedisplayskip  3pt plus 10pt minus 10pt
\global\belowdisplayskip 3pt plus 10pt minus 10pt
\def\rtitle##1{\centerline{\undertext{\twentyfourrm ##1}}}
\def\ititle##1{\centerline{\undertext{\twentyfourit ##1}}}
\def\ctitle##1{\centerline{\undertext{\caps ##1}}}
\def\vstrut{\hbox{\vrule width 0pt height .35in depth .15in }}
\def\cline##1{\centerline{\vstrut ##1}}
\output{\shipout\vbox{\vskip .5in
\pagecontents \vfill
\hbox to \the\hsize{\hfill{\tenbf \label} } }
\global\advance\count0 by 1 }
\rm }


%
%
%

%

%

%

%
{\obeyspaces\global\let =\ }
%
%
%
\widowpenalty 1000
\thickmuskip 4mu plus 4mu
\unlock
%
\Pubnum={${\twelverm IU/NTC}\  \the\pubnum $}
\pubnum={0000}
\def\p@nnlock{\begingroup \tabskip=\hsize minus \hsize
   \baselineskip=1.5\ht\strutbox \topspace-2\baselineskip
   \noindent\strut\the\Pubnum \hfill \the\date   \endgroup}
\def\titlepage{\FRONTPAGE\paperstyle\p@nnlock}
\def\displaylines#1{\displ@y
  \halign{\hbox to\displaywidth{$\hfil\displaystyle##\hfil$}\crcr
    #1\crcr}}
\def\addressee#1{\null
   \bigskip\medskip\rightline{\the\date\hskip 30pt}
   \vskip\lettertopfil
   \ialign to\hsize{\strut ##\hfil\tabskip 0pt plus \hsize \cr #1\crcr}
   \medskip\vskip 3pt\noindent}
\def\tmsaddressee#1#2{
   \vskip\lettertopfil
  \setbox0=\vbox{\singlespace \halign{\tabskip 0pt \strut ##\hfil\cr
   \noalign{\global\dt@ptrue}#1\crcr}}
  \line{\hskip 0.7\hsize minus 0.7\hsize \box0\hfil}
   \bigskip
   \vskip .2in
   \ialign to\hsize{\strut ##\hfil\tabskip 0pt plus \hsize \cr #2\crcr}
   \medskip\vskip 3pt\noindent}
\def\makeheadline{\vbox to 0pt{ \skip@=\topskip
      \advance\skip@ by -12pt \advance\skip@ by -2\normalbaselineskip
      \vskip\skip@  \vss
      }\nointerlineskip}
\def\signed#1{\par \penalty 9000 \bigskip \vskip .06in\dt@pfalse
  \everycr={\noalign{\ifdt@p\vskip\signatureskip\global\dt@pfalse\fi}}
  \setbox0=\vbox{\singlespace \halign{\tabskip 0pt \strut ##\hfil\cr
   \noalign{\global\dt@ptrue}#1\crcr}}
  \line{\hskip 0.5\hsize minus 0.5\hsize \box0\hfil} \medskip }
\def\lettersize{\hsize=6.25in\vsize=8.5in\hoffset=0in\voffset=1in
   \skip\footins=\smallskipamount \multiply\skip\footins by 3 }
%
%
%
%
%
\outer\def\newnewlist#1=#2&#3&#4&#5;{\toks0={#2}\toks1={#3}%
   \dimen1=\hsize  \advance\dimen1 by -#4
   \dimen2=\hsize  \advance\dimen2 by -#5
   \count255=\escapechar \escapechar=-1
   \alloc@0\list\countdef\insc@unt\listcount     \listcount=0
   \edef#1{\par
      \countdef\listcount=\the\allocationnumber
      \advance\listcount by 1
      \parshape=2 #4 \dimen1 #5 \dimen2
      \Textindent{\the\toks0{\listcount}\the\toks1}}
   \expandafter\expandafter\expandafter
    \edef\c@t#1{begin}{\par
      \countdef\listcount=\the\allocationnumber \listcount=1
      \parshape=2 #4 \dimen1 #5 \dimen2
      \Textindent{\the\toks0{\listcount}\the\toks1}}
   \expandafter\expandafter\expandafter
    \edef\c@t#1{con}{\par \parshape=2 #4 \dimen1 #5 \dimen2 \noindent}
   \escapechar=\count255}
\def\c@t#1#2{\csname\string#1#2\endcsname}
%
%
%
%
%
%
%
\def\noparGENITEM#1;{\hangafter=0 \hangindent=#1
    \ignorespaces\noindent}
\outer\def\noparnewitem#1=#2;{\gdef#1{\noparGENITEM #2;}}
\noparnewitem\spoint=1.5\itemsize;
%
%
%
\def\MEMO{\letterstyle\FRONTPAGE \letterfrontheadline={\hfil}
      \hoffset=1in \voffset=1.21in
    \line{\hskip .8in  \special{overlay ntcmemo.dat}
          \quad\fourteenrm NTC MEMORANDUM\hfil\twelverm\the\date\quad}
    \medskip\medskip \memod@f}

\def\memit@m#1{\smallskip \hangafter=0 \hangindent=1in
      \Textindent{\caps #1}}
\def\memod@f{\xdef\to{\memit@m{To:}}\xdef\from{\memit@m{From:}}%
     \xdef\topic{\memit@m{Topic:}}\xdef\subject{\memit@m{Subject:}}%
     \xdef\rule{\bigskip\hrule height 1pt\bigskip}}
\memod@f
\lock
%
%
%
%
\def\NPrefmark#1{\attach{\scriptscriptstyle  #1 }}

%

%
%
%
\def\papersize{\hsize=6.5in\vsize=8.60in\hoffset=0.2in\voffset=2pc
                \skip\footins=\bigskipamount}
\paperstyle
%
%
\doublespace
\pretolerance=500
\tolerance=500
\widowpenalty 2000
\thinmuskip=4mu
\medmuskip=5mu plus 2mu minus 4mu
\thickmuskip=7mu plus 5mu
\PHYSREV
%
%
\REF\NMCG{P. Amaudruz et al., (NMC collaboration),
                   Nucl. Phys. {\bf B371}, 553 (1992);
          M. de Jong, Ph.D thesis, Free University of Amsterdam (1991).}
\REF\LANL{For recent J/$\psi$ production experiments for nuclei, see
          D. M. Alde et al. (E772 collaboration),
                   Phys. Rev. Lett. {\bf 66}, 133 (1991).}
\REF\EMCX{J. J. Aubert et al. (EMC collaboration),
                          Phys. Lett {\bf 123B}, 275 (1983);
          for summaries, see E. L. Berger and F. Coester,
                        Annu. Rev. Nucl. Part. Sci. {\bf 37}, 463 (1987);
             R. P. Bickerstaff and A. W. Thomas,
                        J. Phys. G {\bf 15}, 1523 (1989).}
\REF\KMRS{For recent parametrizations of parton distributions, see
          P. N. Harriman, A. D. Martin, W. J. Stirling,
           and R. G. Roberts, Phys. Lett. {\bf 243B}, 421 (1990);
                              Phys. Rev. {\bf D42}, 798 (1990);
          J. Kwiecinski, A. D. Martin, W. J. Stirling,
           and R. G. Roberts, Phys. Rev. {\bf D42}, 3645 (1990);
          the KMRS-B0 parametrization is used
          for input parton distributions in this investigation.}
\REF\GR{M. Gl\"uck, E. Hoffmann, and E. Reya,
                      Z. Phys. {\bf C13}, 119 (1982).}
\REF\BCDMS{F. Bergsma et al. (CHARM collaboration),
                      Phys. Lett. {\bf 123B}, 269 (1983);
           A. C. Benvenuti et al. (BCDMS collaboration),
                      Phys. Lett. {\bf 223B}, 490 (1989).}
\REF\WASEVEN{M. Bonesini et al. (WA70 collaboration),
                      Z. Phys. {\bf C38}, 371 (1988) and references therein.}
\REF\ABFOW{E. N. Argyres, A. P. Contogouris, N. Mebarki,
           and S. D. P. Vlassopulos, Phys. Rev. {\bf D35}, 1584 (1987);
           P. Aurenche, R. Baier, M. Fontannaz, J. F. Owens, and M. Werlen,
                      Phys. Rev. {\bf D39}, 3275 (1989). }
\REF\NMCGA{D. Allasia et al. (NMC collaboration),
                      Phys. Lett. {\bf 258B}, 493 (1991).}
\REF\KIM{C. S. Kim, Nucl. Phys. {\bf B353}, 87 (1991);
         M. Dress and C. S. Kim, Z. Phys. {\bf C53}, 673 (1992).}
\REF\MQ{ A. H. Mueller and J. Qiu, Nucl. Phys. {\bf B268}, 427 (1986);
                          J. Qiu, Nucl. Phys. {\bf B291}, 746 (1987).}
\REF\CQR{ F. E. Close, J. Qiu, and R. G. Roberts,
                          Phys. Rev. {\bf D40}, 2820 (1989).}
\REF\FF{ L. L. Frankfurt and M. I. Strikman,
                 Phys. Rev. Lett. {\bf 65}, 1725 (1990).}
\REF\KWI{J. Kwiecinski, Z. Phys. {\bf C45}, 461 (1990);
         J. Collins and J. Kiecinski,
                        Nucl. Phys. {\bf B335}, 89 (1990).}
\REF\CERN{R. Albrecht et al. (WA80 collaboration),
                      Phys. Lett. {\bf 201B}, 390 (1988);
                      Z. Phys. {\bf C51}, 1 (1990).}
\REF\FNAL{G. Alverson et al. (E706 collaboration),
                   Phys. Rev. Lett. {\bf 68}, 2584 (1992).}
\REF\BERGER{E. L. Berger and D. Jones, Phys. Rev. {\bf D23}, 1521 (1981);
            A. D. Martin, C. K. Ng, and W. J. Stirling,
                          Phys. Lett. {\bf 191B}, 200 (1987).}
\REF\MATSUI{T. Matsui and H. Satz, Phys. Rev. {\bf 178B}, 416 (1986).}
\REF\SKFEC{S. Kumano and F. E. Close,
               Phys. Rev. {\bf C41}, 1855 (1990).}
\REF\SKA{S. Kumano, Indiana University preprint IU/NTC92-02;
                 talk given at the 13th International Conference
                            on Few Body Problems in Physics,
                            Adelaide, Australia,
                            Jan. 5$-$11, 1992;
              in Proceedings of the International Workshop on
                 Gross Properties
                 of Nuclei and Nuclear Excitations,
                 Hirschegg, Austria, Jan. 20$-$25, 1992,
                 edited by H. Feldmeier;
              to be submitted for publication.}
\REF\RESCALE{F. E. Close, R. G. Roberts, and G. G. Ross,
                          Phys. Lett. {\bf 129B}, 346 (1983);
                          Nucl. Phys. {\bf B296}, 582 (1988);
             F. E. Close, R. L. Jaffe, R. G. Roberts, and G. G. Ross,
                          Phys. Rev. {\bf D31}, 1004 (1985). }
\REF\SKB{S. Kumano, preprint IU/NTC92-20 (1992),
                    submitted to Phys. Rev. C.}
\REF\ANALYSIS{For evaluating the modifications $\Delta G(x)$,
              see comments in the appendix of Ref. \SKB.}
\REF\LLE{C. H. Llewellyn Smith, Nucl. Phys. {\bf A434}, 35c (1985).}
\REF\RALSTON{J. P. Ralston, Phys. Lett. {\bf 172B}, 430 (1986).}
\REF\NA{M. C. Abreu et al. (NA38 collaboration),
                         Z. Phys. {\bf C38}, 117 (1988).}
\REF\BRAUN{P. Braun-Munzinger and G. David,
              in Proceedings of the International Workshop on
                 Gross Properties
                 of Nuclei and Nuclear Excitations,
                 Hirschegg, Austria, Jan. 20$-$25, 1992,
                 edited by H. Feldmeier.}
\FIG\FIGONE{Modifications of $G(x)$ due to parton recombinations.}
\FIG\FIGTWO{Predicted gluon distributions in the nuclei (a) C and (b) Sn.
                The dashed curves indicate parton recombination effects
                (Eq. (2))
                and the solid curves are combined results (Eq. (3))
                of the recombination and the $Q^2$ rescaling.
                $\xi_{_A}$=1.60 for C and 2.24 for Sn.}
\FIG\FIGTHREE{Comparisons of our theoretical results with
                NMC data [\NMCG].
                The dashed curve indicate parton recombination effects
                with $z_0$=0
                and the solid (dash-dot) curve shows combined results
                of the recombination and the $Q^2$ rescaling
                with $z_0$=0 ($z_0$=2 fm). }
\pubnum={92-21}
\pubtype={ }
\date={October 5, 1992}
\titlepage

$~~~$

$~~~$

\title{{\bf Nuclear Gluon Distributions in a Parton Model}}

\vskip 0.6cm

\author{S. Kumano$^*$}

\medskip

\address{Nuclear Theory Center, Indiana University}
\address{2401 Milo B. Sampson Lane}
\address{Bloomington, Indiana 47408-0768, U.S.A.}

\medskip

$~$

$~$

\abstract
Gluon distributions in the carbon and tin nuclei
are investigated by using a $Q^2$ rescaling model with
parton recombination effects.
We obtain strong shadowings in the small $x$ region
due to the recombinations.
The ratio $G_A(x)/G_N(x)$ in the medium $x$ region
is typically 0.9 for medium size nuclei.
At large $x$, the ratio becomes large due to
gluon fusions from different nucleons.
Comparisons with recent New Muon Collaboration data
for $G_{Sn} (x)/ G_C (x)$
indicate that more accurate experimental data
are needed for testing the model.

\vfill
\noindent
\hrule width6cm

\noindent
* present address: Institut f\"ur Kernphysik, Universit\"at Mainz,
6500 Mainz, Germany.

$~~~$

\rightline
{submitted to Phys. Lett. B}
\endpage
\sequentialequations
%

Recent measurements of gluon distribution ratios $G_{Sn}(x)/G_C(x)$
by New Muon Collaboration (NMC) [\NMCG]
are the first data which could shed light on
gluon distributions in nuclei [\LANL].

Modifications of the structure function $F_2(x)$
in nuclei were discovered by
European Muon Collaboration (EMC effect) [\EMCX].
This effect has been an interesting topic in the sense
that it may provide an explicit quark signature
in nuclear phenomena.
On the contrary, ``gluonic EMC effect'' is little known.
Gluon distribution functions in the nucleon
[\KMRS] have been
investigated by using
muon (electron, or neutrino) deep inelastic
scattering data [\GR,\BCDMS],
direct photon data [\WASEVEN,\ABFOW], and muon induced
J/$\psi$ production data [\NMCGA,\KIM].
Although
there are some theoretical predictions [\MQ,\CQR,\FF,\KWI]
for gluon distributions in nuclei,
we have little experimental data.
There are direct photon data for proton
reactions with nuclear targets [\CERN,\FNAL]; however,
there is no available large $p_{_T}$ data in
the WA80 case [\CERN].
Accuracy is not good enough
for extracting a gluon distribution
function of the beryllium (Be) nucleus
from the E706 data [\FNAL].
In any case, we do not expect much modification of the gluon
distribution
in light nuclei such as Be.

The NMC analyzed inelastic J/$\psi$ production data
by a color singlet model [\BERGER] and obtained
gluon distribution ratios $G_{Sn}(x)/G_C(x)$ [\NMCG].
These are interesting data which indicate modifications
of the gluon distribution in nuclei.
We should note that it is very important to know the
gluon distributions in nuclei.
For example, J/$\psi$ suppressions in heavy-ion collisions
were proposed as a signature of quark-gluon plasma [\MATSUI].
Although other initial and final state interactions may explain
the J/$\psi$ suppression phenomena, it is important
that initial conditions as local (gluonic) EMC effects [\SKFEC]
should be subtracted out for investigating
the physics origin of the suppressions [\SKA].
Such issues have not been well studied yet because
there is little data for gluon distributions in nuclei.

As a model for explaining the EMC effect, we take
a $Q^2$ rescaling model [\RESCALE] and apply it for
the gluon distribution. The model was proposed originally
for the structure function $F_2(x)$ by considering
the possibility that an effective confinement radius for quarks
is changed in a nuclear environment.
In fact, strong nucleon overlaps are expected in nuclei
by noting the fact that the nucleon diameter
is approximately equal to the average nucleon separation
in nuclei. From this confinement-radius change and the $Q^2$
evolution equation, the $Q^2$ rescaling model was proposed [\RESCALE].
In this research, we use the above simple picture
also for gluons. Namely, the nuclear gluon distribution function
in the rescaling model
is given by
$G_A(x,Q^2) = G_N (x, \xi_{_A} Q^2)$, where
$\xi_{_A}$ is called as the rescaling parameter.

Other new features which exist if a nucleon resides in a nucleus
are parton recombinations (fusions). The mechanism has been investigated
for explaining the shadowing region $F_2(x<0.1)$
in parton models [\MQ,\CQR,\SKB].
The recombinations also have important effects
on the gluon structure function due to processes
shown in Fig. 1.
We find that there are strong shadowings
in the small $x$ region ($x<0.05$) and
strong anti-shadowings in the medium $x$ region [\CQR].
In Ref. \SKB, it was shown that a model
of the $Q^2$ rescaling with the recombination
can explain experimental data of $F_2(x)$ fairly well
in the wide $x$ region ($0.005<x<0.8$).

In this report, the gluonic EMC effect is investigated
by the $Q^2$ rescaling model combined with
the parton recombinations in Ref. \SKB.
Using this model
we calculate gluon distributions in
the carbon (C) and tin (Sn) nuclei.
Results in our model are compared with the recent
NMC ratios $G_{Sn}(x)/G_C(x)$.

In order to investigate recombination effects, we
calculate contributions to $G(x)$ from processes
in Fig. 1 as investigated by Close, Qiu, and Roberts [\CQR].
For example, a gluon with momentum fraction $x$
is produced by a fusion of gluons from the nucleon 1 and 2.
A modification of a parton distribution
$p_3(x_3)$, due to the
process of producing the parton $p_3$ with the momentum
fraction $x_3$
by a fusion of partons $p_1$ and $p_2$, is given by [\CQR,\SKB]
$$
\Delta p_3 (x_3) = K \int dx_1 dx_2
                 ~p_1 (x_1)~ p_2 (x_2) ~
\Gamma_{p_1 p_2 \rightarrow p_3} (x_1,x_2,x_3=x_1+x_2)
{}~\delta (x,x_1,x_2)~~,
\eqno(1)
$$
where $K$ is given by $K=9A^{1/3}\alpha_s/(2R_0^2 Q^2)$
with $R_0$=1.1 fm
and the strong-interaction coupling constant is
$\alpha_s (Q^2) = 4 \pi/ [9 ~ln (Q^2/\Lambda^2)]$.
The $\delta$ function is given by
$\delta (x-x_1-x_2)$ for the processes in Figs. 1a, 1d, and 1e and
$\delta (x-x_1)$     for Figs. 1b, 1c, 1f, 1g, 1h, and 1i.
The parton fusion function $\Gamma(x_1,x_2,x_3)$
is a probability for producing a parton $p_3$
with momentum fraction $x_3$ by a fusion of partons $p_1$ and $p_2$
with momenta $x_1$ and $x_2$ respectively.

Now, we discuss numerial analysis.
In evaluating recombination contributions $\Delta G(x)$ by
using Eq. (1) [\ANALYSIS], we assume that
a leak-out parton (we denote $p^*(x)$)
is a sea quark or a gluon and that
the momentum cutoff function [\CQR,\LLE]
for this parton is taken as $w(x)=exp(-m_{_N}^2 z_0^2 x^2/2)$
with $z_0$=0 or 2 fm. Then, distributions for the leak-out partons
are $q^*(x)=w(x) q_{sea} (x)$, $\bar q^*(x)=w(x) \bar q (x)$,
and $G^* (x)=w(x) G(x)$.
Input parton distributions
are given by a recent parametrization in Ref. \KMRS.
$Q^2=5~\rm GeV^2$ is used in the parametrization
and for calculating $K$ in Eq. (1).
The QCD scale parameter $\Lambda$ in $\alpha_s (Q^2)$
is taken as $\Lambda$=0.2 GeV.
In our theoretical analysis, targets nuclei are assumed as
$^{12}C$ or $^{118}Sn$.
The rescaling parameters for these nuclei are taken from
Ref. \RESCALE, and they are
$\xi_{_A}(C) $=1.60 and $\xi_{_A}(Sn)$=2.24.

Before discussing the gluon distributions in C and Sn,
we first check that our model can explain
structure functions $F_2^A(x)$ of these nuclei.
As shown in Ref. \SKB, our results are consistent
with SLAC, EMC, and E665 experimental data
for $F_2^A(x)$ of C, Ag, Sn, and Xe nuclei
in the wide $x$ range ($0.005<x<0.8$).
In explaining these data, the most important factor
is the gluon shadowing.
Taking modified parton distributions
due to the recombinations at small $Q^2$,
we should calculate distributions at large $Q^2$,
where the structure functions were measured.
Instead of solving the evolution exactly,
we simply used a solution  [\RALSTON]
for the Altarelli-Parisi equation in the small $x$ region
($\displaystyle{
x \delta q^{sea}_i (x)= - { x \over {12}}
                    {\partial \over {\partial x}}
                    [x \Delta G(x)]
}$, where i=u, d, or s).
This approximate way of treating the evolution
violates the momentum conservation even though
it is satisfied in the recombinations.
The effect due to this extra term is given by
$\displaystyle{
6 \int dx ~x~ \delta q^{sea}_i (x)
}\displaystyle{
=+{1 \over 2} \lim_{x\rightarrow 0} x^2 \Delta G(x)
 +{1 \over 2} \int dx x \Delta G(x)
}$.
The first term vanishes if the input gluon distribution
satisfies, for example,
$\displaystyle{\lim_{x\rightarrow 0} x G(x)=}$constant.
Dominant contributions to $\Delta G(x)$ come from
the gluon-gluon fusion processes; however,
they satisfy
the momentum conservation by themselves
($\int dx x \Delta G_{GG\rightarrow G}(x)=0$).
Therefore, the violation is a small effect due
to $q\bar q\rightarrow G$, $Gq\rightarrow q$, and
$G \bar q \rightarrow \bar q$ processes.
A numerical evaluation for the Ca nucleus indicates
that such violation effect is less than 1\%
(6$\int dx x \delta q^{sea}_i (x)=-0.005$); hence
it is not a serious effect on the momentum conservation.

Using our model, which can explain at least
the structure function $F_2^A(x)$, we predict
$G_C(x)$ and $G_{Sn}(x)$.
We take the same rescaling parameter for
all partons by taking a naive consideration
that the confinement radius change modifies all parton
momenta at the same rate.
Predicted gluon distributions for C and Sn are
shown in Figs. 2a and 2b. In these figures, the dashed curves
are recombination effects shown by
$$
R_A ({\rm recombination}) ~=~ 1~+~
{{ \Delta G_A (x, Q^2) } \over { G_N (x,Q^2) } }  ~~~,
\eqno(2)
$$
$\Delta G_A(x,Q^2)$ is calculated by using Eq. (1)
(note that no $Q^2$ rescaling is used)
and explicit expressions are given in Ref. \SKB.
Solid curves are combined contributions
from the rescaling and recombinations and they
are shown by the ratio
$$
\displaystyle{
{{G_A (x,Q^2)} \over {G_N (x,Q^2)}} ~=~
{{\tilde G_A (x,Q^2) + \Delta \tilde G_A (x, Q^2)} \over
 { G_N    (x,Q^2) } }        ~~~.
}
\eqno(3)
$$
\noindent
In these equations,
$\tilde G_A (x,Q^2)$ and $\Delta \tilde G_A (x,Q^2)$
are given by the rescaling model,
$\tilde G_A (x,Q^2)$

\noindent
=$G_N (x,\xi_{_A} Q^2)$ and
$\Delta \tilde G_A (x,Q^2)=\Delta G (x, \xi_{_A} Q^2)$.
The recombination mechanism produces strong shadowing effects
in the small $x$ region due to gluon-gluon and gluon-quark
fusion processes in Fig. 1.
In the medium-large $x$ region,
contributions are dominated by the gluon-gluon fusion process
in Fig. 1a.
It is interesting to note in our model that
gluon distributions at $x>1$ could be produced
in the fusion process.
This is the reason why
the ratio goes to infinity at $x \rightarrow 1$
in Figs. 2a and 2b.
The $Q^2$ rescaling contributions are opposite to
the recombination. The rescaling effects are positive
in the small $x$ region and are negative in
the medium-large $x$ region.
Combined contributions shown by the solid curves in
Figs. 2a and 2b indicate strong shadowings in
the very small $x$ $(x<0.02)$ region, depletions (0.8$-$1.0)
in the medium $x$ $(0.2<x<0.6)$, and large ratios in
the large $x$ $(x>0.7)$.

In investigating shadowings in $F_2(x)$ in Ref. \SKB,
we used the rescaling for parton distributions at $x<0.1$.
Although the rescaling produces large positive contributions,
they are counterbalanced by the large shadowings produced
through gluon distributions ($\delta F_2$ in Ref. \SKB).
Therefore, combined contributions are not very dependent
(about 5\% differences)
whether or not the rescaling is used in the region ($0.005<x<0.1$).
On the contrary, gluon distributions at small $x$ are
very sensitive to whether or not the rescaling is used
as shown in Figs. 2a and 2b.
If there is no rescaling at $x<0.1$, we should have
strong gluon shadowings in the region $x<0.1$
as shown by the dashed curves.
However, if the rescaling is used, gluon distributions
are shadowed only in the very small $x$ region ($x<0.02$)
as shown by the solid curves.

The nuclear gluon distributions in the medium-large
$x$ region are very sensitive to the momentum cutoff
as shown in Figs. 2a and 2b.
For example, $G_{Sn}(x)/G_N(x)= 1.02$ (for $z_0=0$) at $x=0.4$,
but it is 0.87 if $z_0=2$ fm.
Because $G_N(x)$ itself is very small at $x=0.4$,
it may seem to be an insignificant problem.
However, it is important for describing
e.g. $p_{_T}$ dependence of $J/\psi$
in heavy-ion collisions.
The rapid increase of $G_A(x)/G_N(x)$ in
the medium $x$ region could be responsible for
the $p_{_T}$ dependence of $J/\psi$ suppressions
observed by NA38 [\NA], although
the $p_{_T}$ slope obtained in the local
gluonic EMC effect is rather small [\SKA]
compared with the NA38 data at this stage.
The gluon distributions at the medium $x$
are so sensitive to the momentum cutoff $w(x)$
that we need to study more about the cutoff [\LLE].
We leave the problem of the cutoff
as a future research topic.

Calculated gluon distributions are compared
with the NMC data for $G_{Sn}(x)/$
$G_C(x)$ [\NMCG] in Fig. 3.
The dashed curve shows recombination
results with $z_0$=0 and the solid (dash-dot)
curve shows combined results of the $Q^2$ rescaling
and the recombinations with $z_0$=0 ($z_0=2$ fm).
Our theoretical results shown by the solid and dash-dot
curves indicate ratios $G_{Sn}(x)/G_C(x)<1$ at
$x<0.03$ due to the gluon shadowing. The ratios
are about 1.03 in the region ($0.05<x<0.20$) and
are dependent on the momentum cutoff $w(x)$
in the medium $x~(x>0.2)$.
We notice that experimental errors are very large
in comparison with typical theoretical modifications.
Because our model predictions for the modifications
are less than 5\% in the range $0.05<x<0.2$,
we need accurate data better than 5\% accuracy
in order to test the model.
Because the NMC data for
$G_{Sn} (x) / G_C (x)$ are not accurate enough,
we should wait for better measurements of $G_A(x)$,
for example a proposed experiment at RHIC [\BRAUN],
in the small $x$ region for investigating details
of the gluon shadowing.
It is also important to know gluon distributions
in the medium $x$ region ($0.3<x<0.6$), although
distributions themselves are very small, for
studying J/$\psi$ productions
in heavy-ion collisions [\SKA].

In summary, we investigated
gluon distributions in the nuclei C and Sn
by the rescaling model
with parton recombinations effects.
We obtained shadowings in the nuclear gluon distributions
in the small $x$ region due to the recombinations and
depletions [typically $G_A(x)/G_N(x)\sim 0.9$]
in the medium $x$ region.
The ratio $G_A(x)/G_N(x)$ becomes large at $x>0.6$
due to gluon fusions from different nucleons.
The ratio in the medium-large $x$ region is very
sensitive to the momentum cutoff for leak-out partons
in our model.
Comparisons
with the NMC data indicate that more accurate
experimental data are needed for testing the model.

$~~~~~$

This research was supported by the NSF under
Contract No. NSF-PHY91-08036.
S.K. thanks Drs. F. E. Close, J. Qiu, and R. G. Roberts
for discussions and suggestions about parton recombinations
and direct photon experiments;
Drs. G. K. Mallot and G. van Middelkoop
for discussions about the J/$\psi$ production
experiment by NMC; Drs. T. C. Awes and S. P. Sorensen
for communications about WA80 and RHIC experiments.

\endpage
\par \penalty-400 \vskip\chapterskip
   \spacecheck\referenceminspace \immediate\closeout\referencewrite
   \referenceopenfalse
   \line{\fourteenrm\hfil References\hfil}\vskip\headskip
   \input referenc.tex
   
\endpage
\par \penalty-400 \vskip\chapterskip
   \spacecheck\referenceminspace
   \immediate\closeout\figurewrite \figureopenfalse
   \line{\fourteenrm\hfil Figure Captions\hfil}\vskip\headskip
   \input figures.aux
   
\bye
\end